# Design of Optically Path Length Matched, Three-Dimensional Photonic Circuits Comprising Uniquely Routed Waveguides

**Ned Charles,[1,*] Nemanja Jovanovic,[2,3,4] Simon Gross,[2,5] Paul Stewart,[1] Barnaby Norris,[1] John O'Byrne,[1] Jon S. Lawrence,[2,3,4], Michael J. Withford[2,3,5] and Peter G. Tuthill,[1,5]**

*[1]Sydney Institute for Astronomy (SIFA), School of Physics, University of Sydney, 2006, Australia*
*[2]MQ Photonics Research Centre, Dept. of Physics and Astronomy,*
*Macquarie University, NSW 2109, Australia*
*[3]Macquarie University Research Centre for Astronomy, Astrophysics and Astrophotonics,*
*[4]Australian Astronomical Observatory (AAO), Epping NSW 1710, Australia*
*[5]Centre for Ultrahigh-bandwidth Devices for Optical Systems (CUDOS)*
*[*]Corresponding author: ccharles@physics.usyd.edu.au*

Abstract: A method for designing physically path length matched, three-dimensional photonic circuits is described. We focus specifically on the case where all the waveguides are uniquely routed from the input to output; a problem which has not been addressed to date and allows for the waveguides to be used in interferometric measurements. Circuit elements were fabricated via the femtosecond laser direct-write technique. We demonstrate via interferometric methods that the fabricated circuits were indeed optically path length matched to within 45 μm which is within the coherence length required for many applications.

*OCIS codes: 110.3175, 130.3120, 220.2740, 230.7370, 320.7160*

## 1. Introduction

Integrated photonic circuits stand poised to make a major contribution to a multitude of areas of fundamental science. A recent illustration of their potential comes from the field of quantum optics over the past four years [1]. This technology has been used to realize compact beam combiners (several square millimeters in size) that offer unprecedented levels of stability as compared to their bulk optics counterparts, which will enable large-scale quantum computing in the future.

The inherent advantages offered by integrated photonic technologies have also attracted intense interest within the field of optical stellar interferometry. Indeed, it was almost 20 years ago that starlight from two telescopes was first combined interferometrically with the use of a photonic beam combiner [2]. In this case the astronomers were exploiting an additional advantage of single-mode guided optics commonly referred to as spatial filtering [3]. This property refers to atmospherically perturbed wavefronts from the stellar target being flattened when the light is injected into the fundamental mode of the waveguide. Although this "mode cleaning" is inevitably accompanied by coupling loss, the fact that the two beams have flat wavefronts that are spatially constant results in high signal-to-noise ratio measurements of the fringe visibilities [3]. It is this additional spatial filtering property that has motivated advanced developments of beam combining technologies over the past 20 years in this field [4,5].

The key requirement for any type of interferometric beam combiner is that the paths which the light takes through the interferometer must be matched in length to within the coherence length of the light. So far, this has been achieved in both 2D and 3D integrated platforms through waveguide route symmetry (i.e. the waveguides are simply mirror images of one another), with fine-tuning enabled by means of an optical delay line [6,7]. However, there are emerging applications that place specific constraints on the location of the input/output positions of the waveguides on the end facets of the chip such that each waveguide must take a unique route through the device. One such application is pupil remapping which, in some cases, requires the waveguides to be remapped in three dimensions from a 2D array at the input to a linear array at the output [8,9].

Although complex 3D circuit designs with unique start/end points have been proposed in the past [10], they have not incorporated path length matching. In order to match the optical paths through all arms of the device, an iterative mathematical routine, based on adjusting the curvature and direction of each route individually, has been developed. To date few such algorithms have been described in the literature, with some of the most developed work appearing in the field of machine vision where they were used in a completely different context for image/shape recognition [11-13]. Nevertheless, these concepts spring from the same foundation as the path length matching algorithm for photonic devices presented here.

In this paper, we present the first experimental campaign aimed at designing optically path length matched photonic circuitry which consists of uniquely routed waveguides in three dimensions. Considerable interplay between design, fabrication and testing was required over several generations of devices in order to understand the key limitations to performance and develop strategies to mitigate losses. In section 2, we discuss the techniques and constraints for designing path length matched waveguides. This section demonstrates that when all the constraints are considered in unison, the design process becomes complex in nature. In section 3, we consider the bend and transition losses and develop techniques for predicting the throughputs of the waveguides. In section 4, we look at optimizing the throughput of the guides. Section 5 describes the design and predicted performance of an optimized pupil-remapping device, which consists of eight waveguides. Section 6 summarizes the laboratory results for the performance of an eight waveguide pupil-remapper fabricated by ultrafast laser inscription (ULI).

## 2. Waveguide Curve Design and Path Length Matching

In order to design efficient, path length matched circuits, it is important to first determine the constraints of the system. As our end goal was to design pupil-remapping devices, we were constrained by the requirements for this specific science application. Firstly, in our application, a pupil-remapping device requires that the waveguides are located along the nodal points of a hexagonal lattice at the input of the chip so that a micro-lens array can be used to inject starlight into each waveguide. The waveguides are remapped into a linear array with an equidistant spacing across the output of the device. It is important that the waveguides intersect the output facet at right angles in order to ensure maximum coupling. Finally in order to maximize the throughputs the guides must be continuous with minimal curvature along their entire lengths. Our first-pass solution to generate the waveguide trajectories relied on cubic splines, as illustrated in blue in Fig. 1.

The spline curve was created by initially defining three points along the device, such as those indicated by the dots on the solid blue line in Fig. 1.

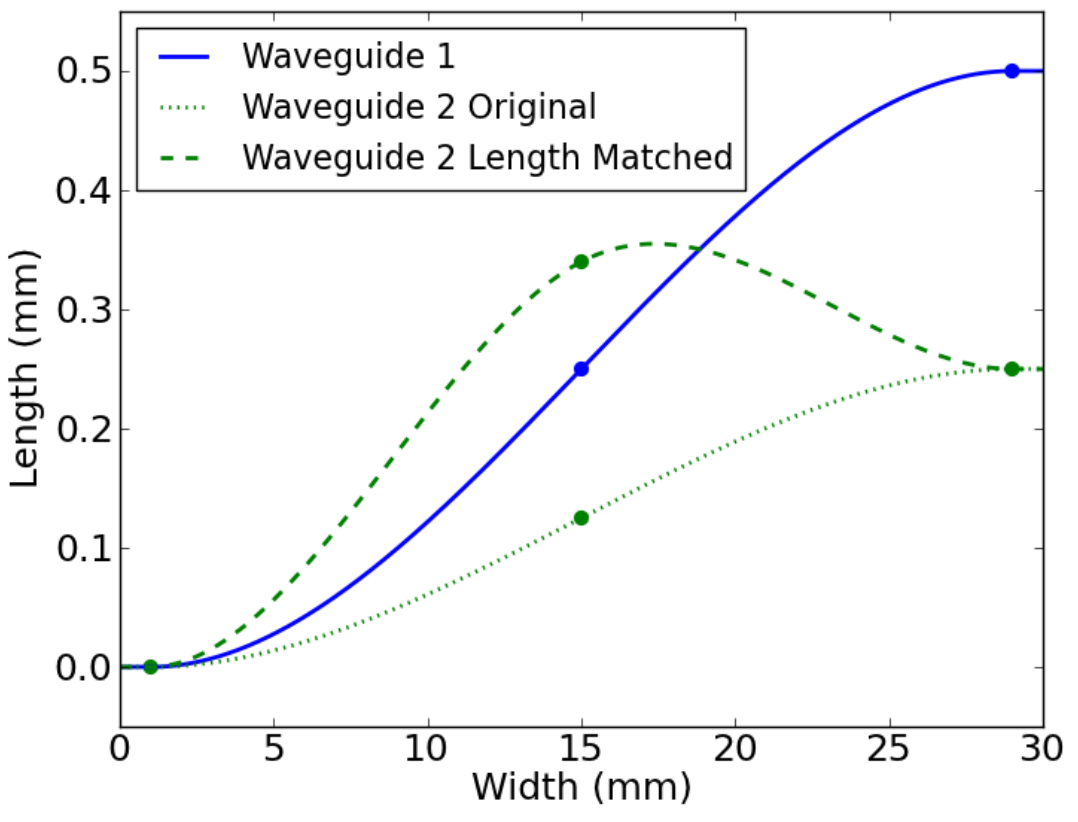

**Fig. 1 – Original cubic spline based waveguide (solid blue line). A second waveguide with a different output coordinate (dotted green line), and its path length matched equivalent (dashed green line).**

First, the endpoints were fixed to the required input/output coordinates and then the route could be adjusted by moving the central point within a mid-plane that was halfway between the two ends. This is a relatively easy concept to extrapolate to three dimensions. An interpolating cubic spline algorithm [14,15] was used to generate the curves between the user-defined points. In the devices illustrated here, 1 mm long straight sections of waveguide were used at either end of the device in order to ensure orthogonality of the guides to the end facets. A second curve (dotted green line) is shown in Fig. 1 with an offset in width, between the input and output, half of that of Waveguide 1. As Waveguide 2 shares the same start point but has half the width offset of Waveguide 1, then Waveguide 2 is shorter and hence the path lengths are not matched.

In order to match the path lengths of the waveguides, the midpoint of the second curve was shifted up in small increments, as shown by the dashed green line in Fig. 1. At each increment the path length was recalculated. The procedure was repeated until path length matching to within 0.1 µm was achieved between the waveguides. In the case where there are many waveguides to be path length matched (our first generation pupil remapping device employed eight guides), then the waveguide with the largest lateral offset between input and output should be used to set the path length for all of the other guides to match. Matching waveguide routes physically compensates for the majority of the path length mismatch as optical effects, such as inhomogeneities in waveguides, are minor in comparison.

Using a two dimensional path length matched spline curve, such as those in Fig. 1, it is relatively simple to create a three-dimensional route for the waveguide. Firstly, we take the coordinate of the endpoint within the output plane and project it back onto the input plane, as shown in Fig. 2, to determine the lateral offset of the spline curve which is given by the length of *r*. This offset is equivalent to the lateral distance between the green dots at the input and output of the waveguide routes shown in Fig. 1. Therefore, by designing a two-dimensional route with a lateral offset that matches the length of *r*, and simply rotating it about the z-axis, it is possible to realize a curve in three-dimensional cylindrical coordinates, which can then be translated into Cartesian coordinates.

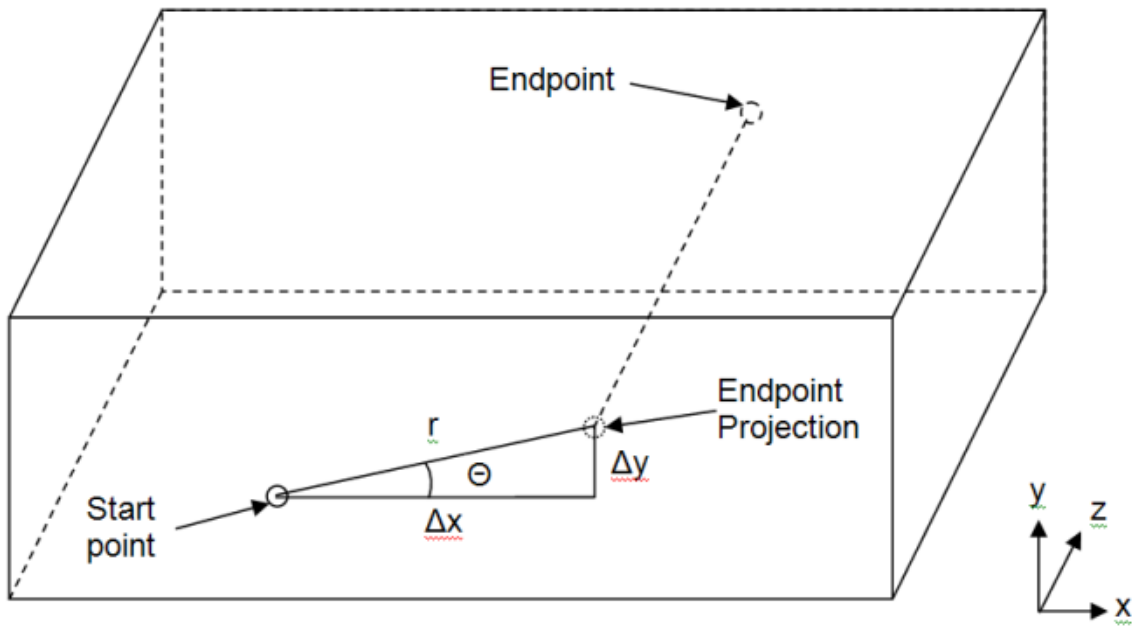

**Fig. 2 – Three-dimensional plot showing start and end points, along with endpoint projection and values needed for conversion from cylindrical to Cartesian coordinates.**

When fabricating the waveguides in three-dimensional space, additional design constraints must also be considered. For example, the waveguides must be spatially separated sufficiently so as to minimize evanescent cross-coupling which is achieved by maintaining a minimum proximity of 30 µm between any two guides. In addition there are several constraints brought about by the fabrication technique itself. For the case of the pupil remappers discussed above, ultrafast laser inscription (ULI) was the fabrication technique of choice owing to its applicability to three-dimensional architectures [16]. One unique consideration of this fabrication technique, in the absence of an adaptive optics system [17], is the limited depth range over which the device can be fabricated due to distortions of the focal spot size induced by the air/glass-interface [18]. An oil immersion objective lens can be used to mitigate the influence of spherical aberration. However the physical working distance of such objectives is typically limited to a few hundred microns (450 µm for the objective used in this work). The second consideration imposed by ULI is that the laser cannot be focused through a previously modified region of substrate without aberration. This would cause the focal volume to be modified and hence the waveguide properties to be altered locally, reducing waveguide performance. This translates into a requirement that all waveguide crossings, viewed from above, could not be so entangled as to pass both over and under at separate places for any pair of guides. This also meant that the waveguides were fabricated in sequence from the deepest to the shallowest.

To satisfy these additional constraints, a new design variable had to be introduced. This new variable allowed for an extra component of rotation in three-dimensions about an axis formed by the two endpoints. By adjusting the rotation angle of the center point in space, and then path length matching the waveguide at that position, it was possible to satisfy all of the above criteria. The effect of performing this extra rotation can be illustrated by inspecting the three-dimensional path of a waveguide as a series of two-dimensional projections onto the x-z and y-z planes as depicted in Fig. 3.

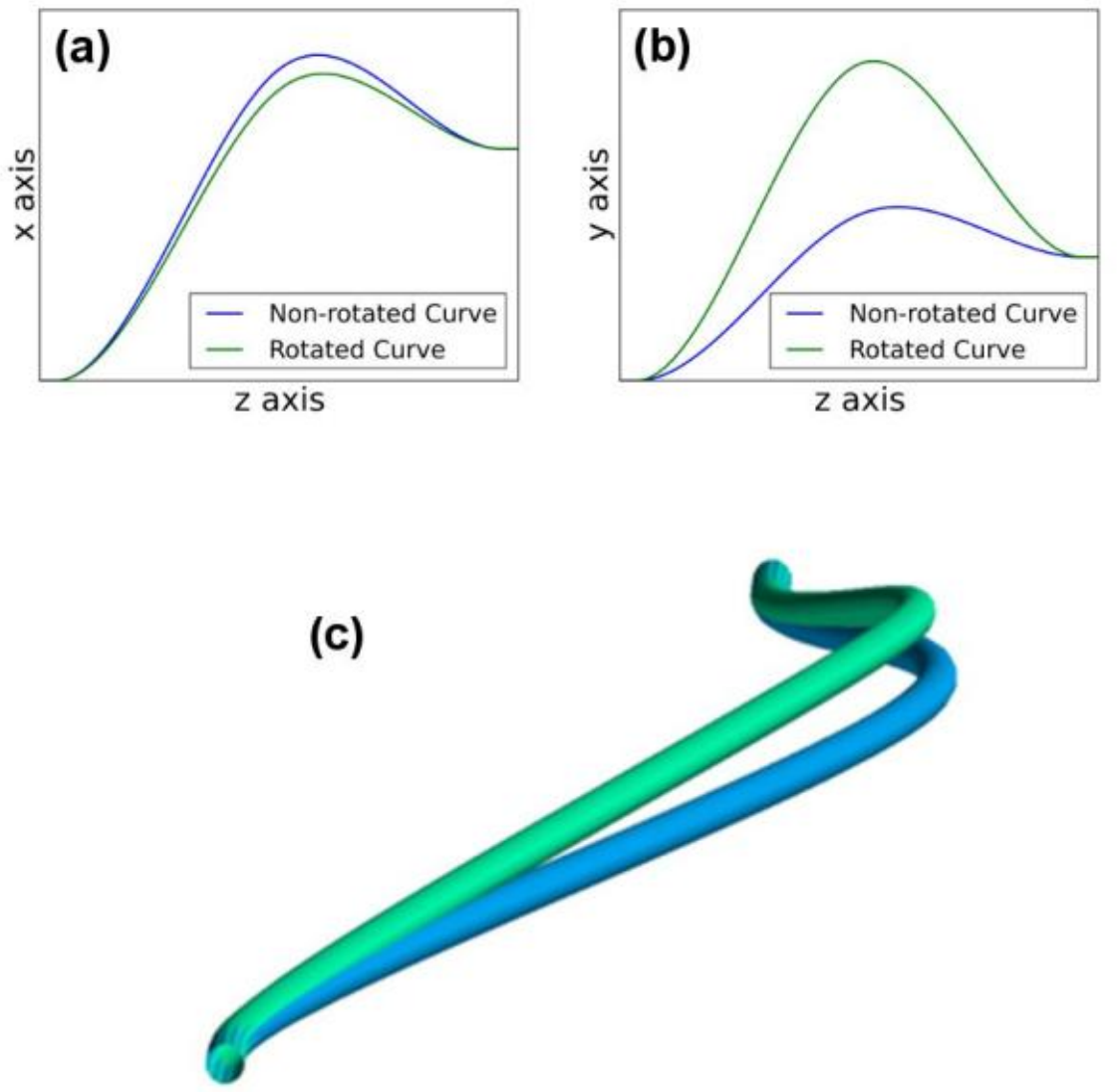

**Fig. 3 – A three-dimensional waveguide before and after rotation about the axis formed by the endpoints (c) shows the effect of the rotation in three-dimensions from a viewpoint like that depicted in Fig. 2. (a) and (b) show the two-dimensional projections in the x-z and y-z planes respectively.**

By employing the techniques outlined above it was found to be possible, with some care and effort, to position multiple waveguides in three-dimensional space within the limited depth range while maintaining the minimum proximity between waveguides. One such pupil-remapping device that consists of eight waveguides is shown in Fig. 4. These waveguides were designed to be physically path length matched to within 0.1 µm.

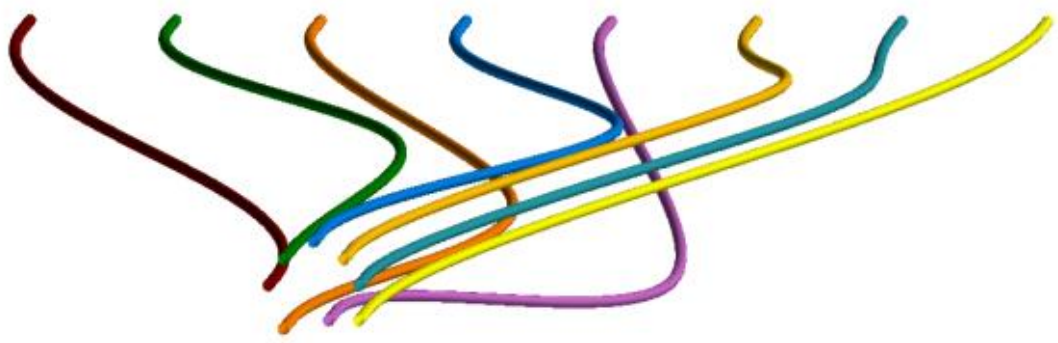

**Fig. 4 – Three-dimensional image of eight path length matched waveguides.**

## 3. Curvature and Power Loss

The curvature of a waveguide at any point can be characterized by an osculating circle that best fits the curve locally. The radius of this osculating circle is known as the radius of curvature. This radius can be used to calculate the bend loss in a waveguide by calculating the difference between input power and output power using the following equation [19],

$$P(z) = P(0)\exp(-\gamma z) \qquad (1)$$

where $z$ is the position along the waveguide from the input and $\gamma$, the bend loss coefficient. The bend loss coefficient for a step-index profile is given by the following equation,

$$\gamma = \frac{\sqrt{\pi}}{2\rho}\left(\frac{\rho}{R_c}\right)^{1/2}\frac{U^2}{V^2W^{1.5}}\frac{1}{K_1^2(W)}\exp\left(\frac{-4}{3}\frac{R_c}{\rho}\frac{W^3D}{V^2}\right) \qquad (2)$$

where $\rho$ is the radius of the core of the waveguide, $R_c$ the radius of curvature, $V$ the normalized frequency, and $U$ and $W$ the bulk refractive index parameters given by

$$U = \rho(k^2n_{co}^2 - \beta^2)^{1/2} \qquad (3)$$

$$W = \rho(\beta^2 - k^2n_{cl}^2)^{1/2} \qquad (4)$$

where $\beta$ is the waveguide propagation constant, and $n_{co}$ and $n_{cl}$ are the indices of refraction for the core and bulk refractive index of the waveguide respectively. $D$ is based on these indices of refraction and given by:

$$D = \frac{n_{co}^2 - n_{cl}^2}{2n_{co}^2} \qquad (5)$$

A combination of equations 1 and 2 reveals that the transmitted power through a curved waveguide is related to the radius of curvature by a double exponential function. This is critical as small changes in radius of curvature can cause very large changes in bend loss. Furthermore, bend loss is also influenced by the index contrast ($\Delta n$) between the core and bulk refractive index.

The normalized throughputs as computed from equations 1 and 2, for a 30 mm long waveguide of constant radius of curvature, as a function of the radius of curvature for three values of *Δn* and the results are shown in Fig. 5. For comparison, these waveguides were also modeled in the commercial beam propagation software, BeamPROP (RSoft) [20]. Using its simulated bend transformation that simplifies the calculations, the waveguides were tested over the same radii of curvature for the same three values of *Δn*. For these simulations, the waveguide core radius was set to 4.85 µm, bulk refractive index to 1.4877 and the wavelength to 1550 nm. The results are also shown in Fig. 5.

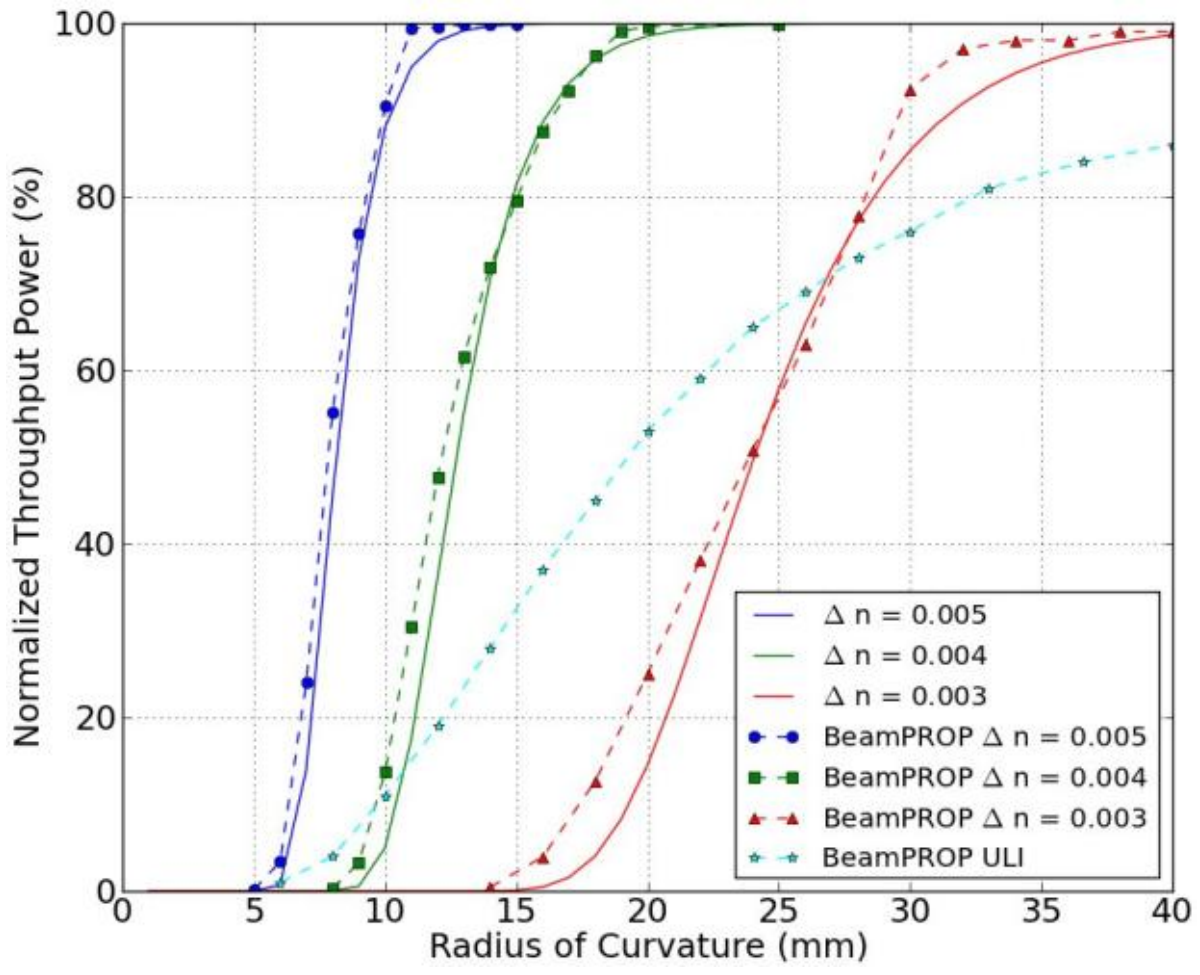

**Fig. 5 – Normalized throughput as a function of radius of curvature for a 30 mm long waveguide. Solid lines are theoretical computations from the equations given while dashed lines are simulations using the BeamPROP package. A step-index profile has been assumed for all guides, except for the Direct Write waveguide (cyan) as described in the text.**

It can be seen that there is a strong correlation between the throughput predictions using the beam propagation method and the analytical expressions for a 30 mm long bend. This demonstrates that either technique could be used for predicting device performance. In addition, the graph illustrates that for each *Δn*, there is a narrow range of radii of curvature over which the throughput changes suddenly with decreasing bend radius (from ~100% down to 0).

Since the index profile of the waveguides fabricated using ULI is not a simple step index profile as modeled in both the techniques described, it was important to ascertain the performance of such a waveguide prior to use. The refractive index profile of a typical waveguide fabricated in Eagle2000 glass in the cumulative heating regime [21] is shown in Fig. 6(a). The refractive index profile was measured at 633 nm with a refracted, near-field profilometer (Rinck Elektronik), and the data inserted into BeamPROP. The index profile was rescaled to compensate for material dispersion between 633 nm (where the index profile was measured) and 1550 nm (where the guides operate) by using the Sellmeier equation [22] for Eagle2000. Finally the index profile was rescaled by a constant factor as the profilometer overestimates the absolute index. The scaling factor was adjusted until the mode field diameter at 1550 nm matched that which was measured in the laboratory. The throughput of 30 mm long bends with the measured laser written index profile is shown in Fig. 5 as a function of the radius of curvature. It is clear that the modeled throughput as a function of radius of curvature is not as steep for the ULI waveguide as it is for the step index profile, which results in waveguides that have non-zero bend losses even when the radii of curvature are large.

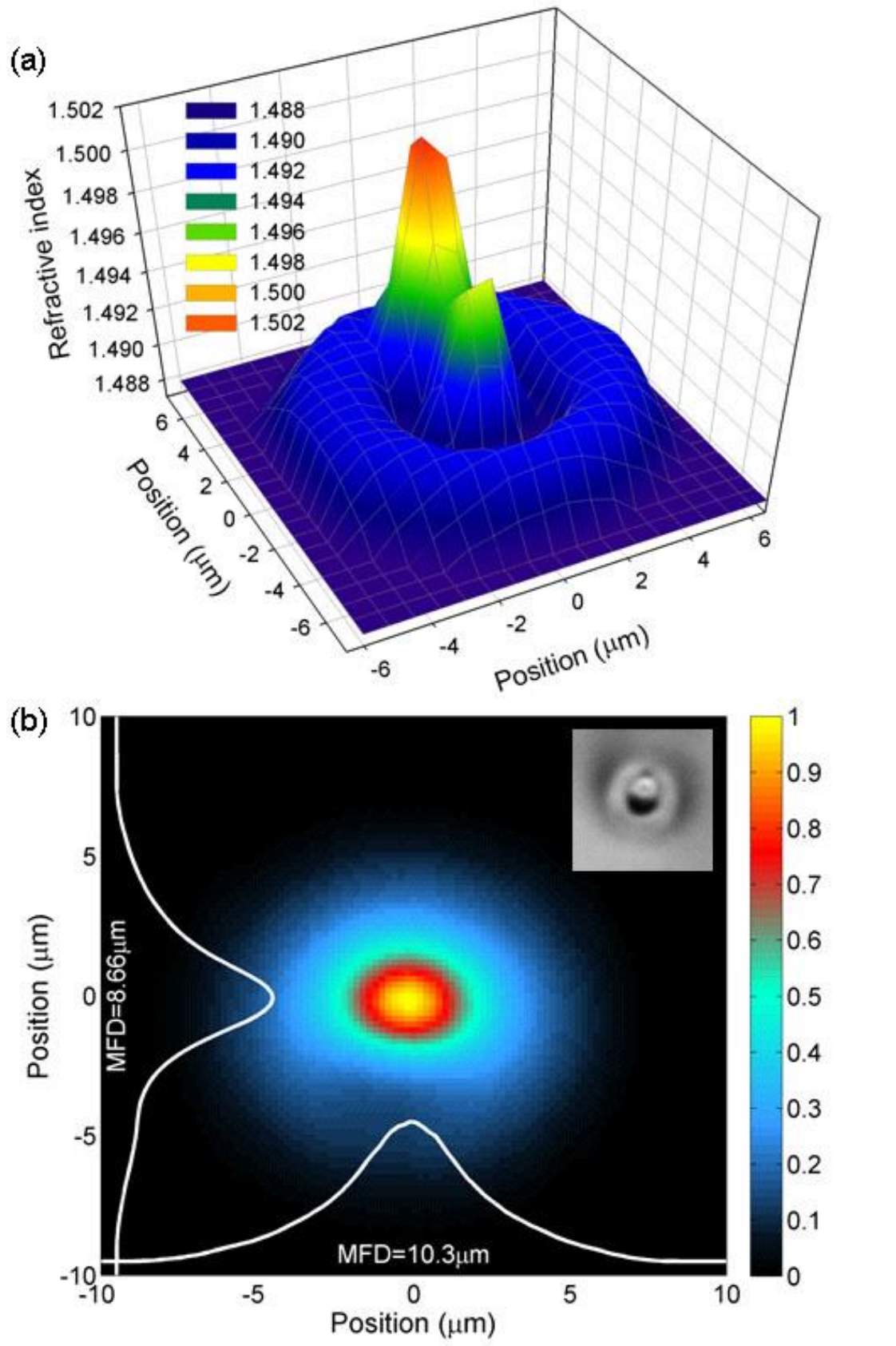

**Fig. 6 – (a) Refractive index profile of an ultrafast laser inscribed waveguide at 1550 nm. (b) Intensity distribution of the mode guided by the waveguide as modeled by BeamPROP. Inset shows a cross-sectional micrograph of the waveguide at the same scale.**

To confirm the behavior predicted in the simulations, a prototype chip consisting of a series of concentric 90° circular arc waveguides of varying radii of curvature was created and the throughputs measured at 1550 nm by butt-coupling a single-mode fiber to the input and output of each waveguide. The transmitted power was normalized against the fiber-to-fiber throughput by directly butt-coupling the two fibers against each other. The bend losses were isolated by removing the affects of coupling and bulk material absorption losses. Loss due to the coupling of light into and out of the chip was determined to be 9 ± 2%, with absorption loss in the Eagle2000 glass at 1550 nm also accounted for ($\alpha$ = 0.0075 ± 0.0003 $mm^{-1}$) [23]. Although the series of guides had varying lengths, the bend losses were rescaled to a waveguide length of 30 mm for comparative purposes. These waveguides were also modeled in RSoft, producing the graph shown in Fig. 7.

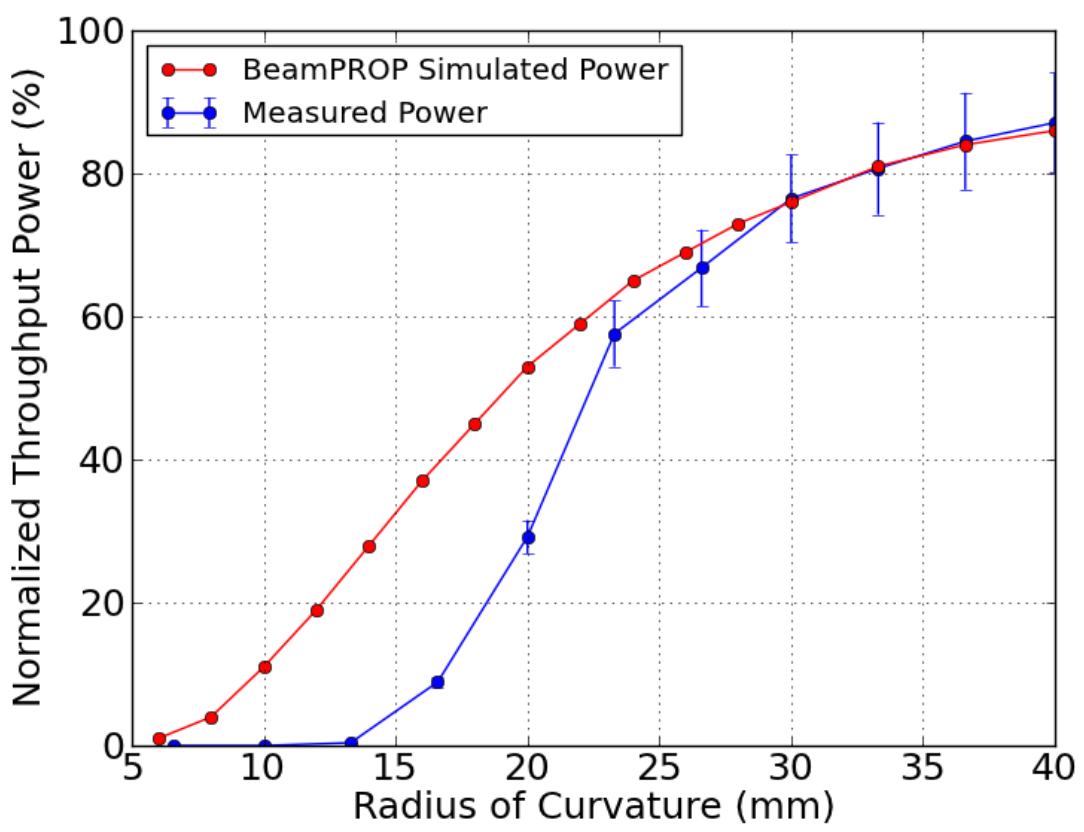

**Fig. 7 – Measured (blue line) and simulated (red line) throughputs for ULI waveguides as a function of radius of curvature.**

The measured and simulated results agree to within the experimental uncertainty for radii above 23 mm, below which the measured throughput drops off more rapidly. This excessive loss for high curvature guides is currently unexplained. However, it is clear that for high efficiency devices, the radii of curvature should be as large as possible and certainly larger than 23 mm, the region where the model agrees with the empirical data.

The final loss mechanism to be considered is the transition loss. A transition loss occurs when light propagates between any two waveguide sections where the radii of curvature differ in magnitude and/or direction. The equation which describes the transition loss is given below [19]:

$$P_{out} = P_{in}\left(1 - \left(\frac{R_1 \pm R_2}{R_1 R_2}\right)^2 \frac{\rho^2 V^4}{8\Delta^2}\left(\frac{r_0}{\rho}\right)^6\right) \qquad (6)$$

It can be calculated from the radii of the two sections ($R_1$ and $R_2$). If the bends are in the same direction, the ± becomes a −. If they are in opposite directions, it becomes a +. If the transition is between a straight section and curved section of radius $R_C$, the $(R_1 \pm R_2)/(R_1R_2)$ term is replaced by $1/R_C$. All values for the equation have been given previously, except for $r_0$, which can be calculated by:

$$r_0 = \frac{\rho}{\sqrt{V-1}} \qquad (7)$$

## 4. Optimizing Waveguide Throughput

Although cubic spline curves can be easily implemented in the design methodology outlined above, they are not an ideal choice when it comes to minimizing bend losses which can become catastrophic where bends become very tight. A more appropriate choice is the use of sections of circular arcs. These curves evenly distribute the curvature over the entire length of the guide instead of producing regions of greater bending to a few key points as is the case for the cubic spline functions. To design an arc-based waveguide which fits the same end points as the blue curve in Fig. 1, two circular arc sections were created from two tangential circles with straight end sections as shown in Fig. 8.

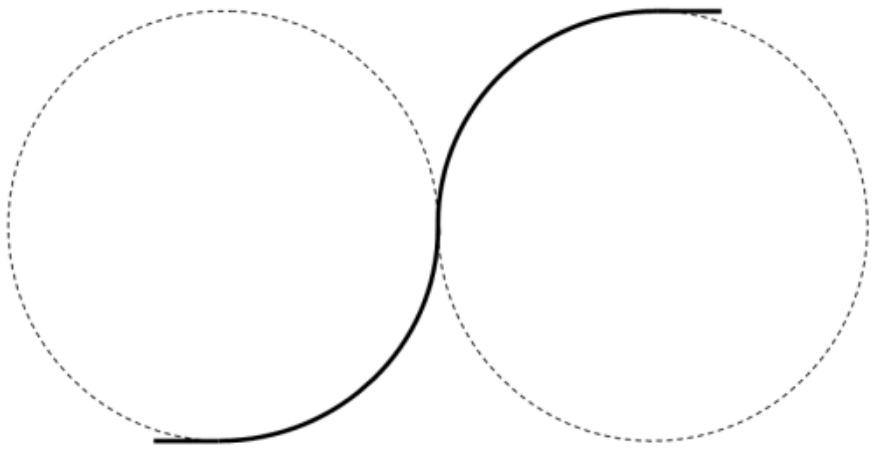

**Fig. 8 – Curve created by two circular arcs.**

Although the creation of a simple waveguide from two circles is relatively straightforward, path length matching is more difficult, as the total path length cannot be increased by using two circular arcs of constant radius alone. Indeed, even if the point of transition between the two arcs was shifted, it can be shown that this does not alter the path length. The addition of a third circle is required, as shown in Fig. 9.

To increase the path length, the radii of all circles must be decreased, with circles 1 and 3 positioned such that the bottom of each circle is at an end point of the waveguide. The middle circle is then positioned at single points of contact with the other two, and the waveguide constructed from arc sections from all three circles, transitioning at the contact points. If we allow the radii of all three circles to decrease equally, the circle labelled 2 will move up and to the left in

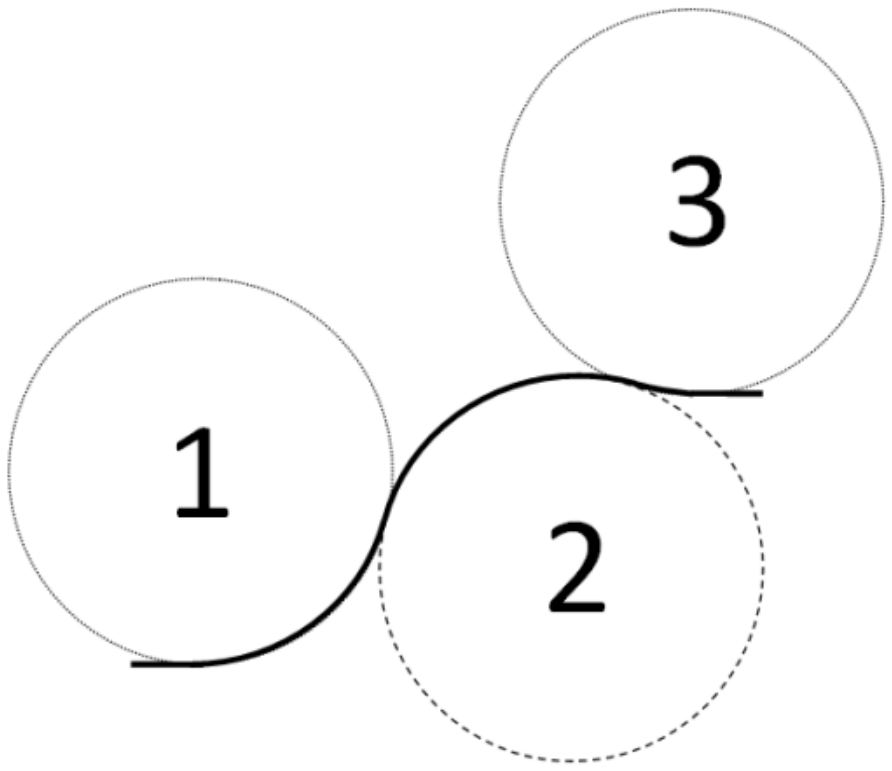

**Fig. 9 – Addition of a third circular arc with reduced radii of all circles, allows for a mechanism to increase path length.**

Fig. 9 to stay in contact with the other circles, increasing the path length of the waveguide until it matches the requirement. To meet proximity requirements, the position of the waveguides can also be adjusted in three-dimensional space using two individual arc-based curves in the x-z and y-z planes respectively similar to the process in Fig. 3.

For comparison, two waveguides with common start and end points were designed using the cubic spline based curves as well as the arc-based curves. The radius of curvature as a function of position along the waveguide is shown in Fig. 10 below for both the cubic spline (green line) and arc-based (blue) designs.

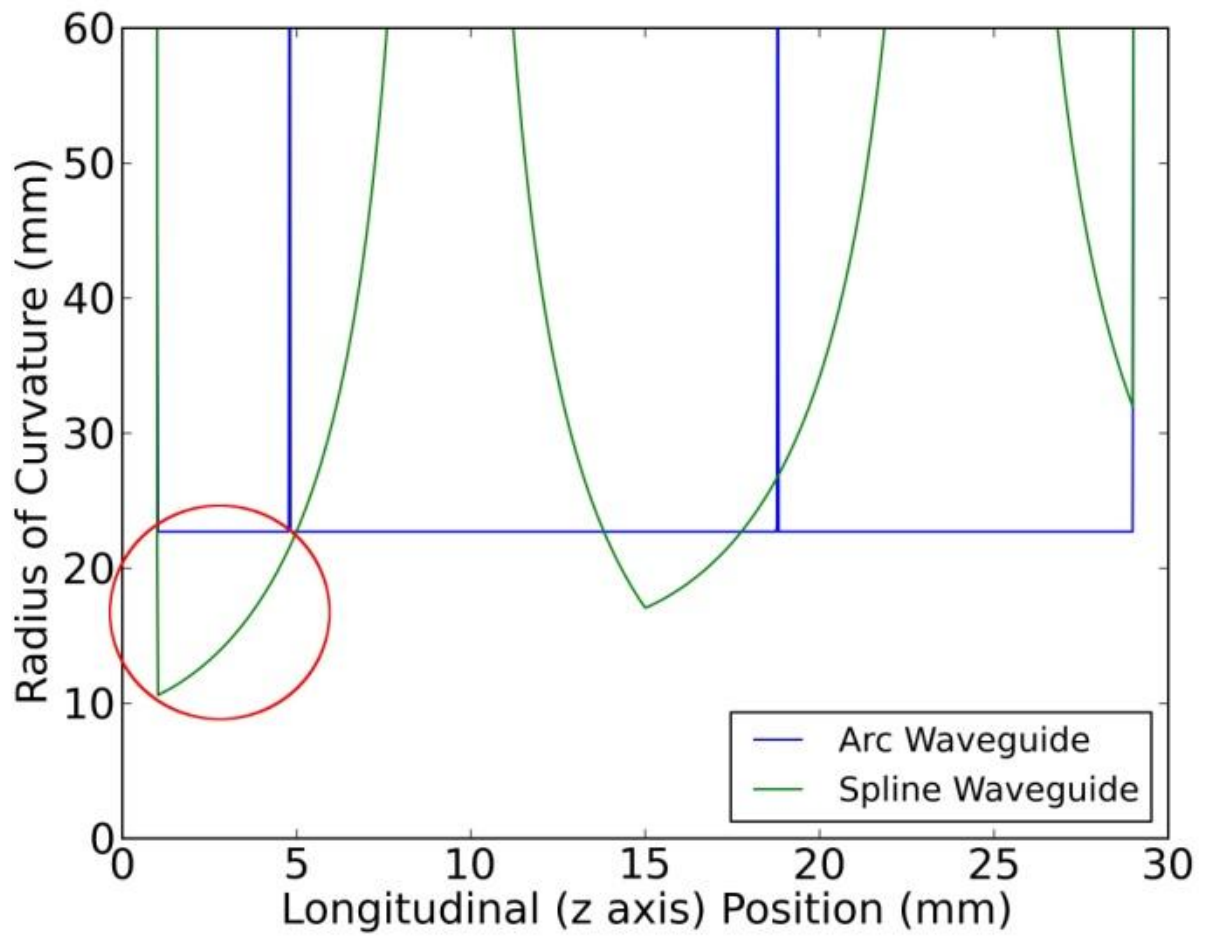

**Fig. 10 – Radius of curvature as a function of position along the waveguide routes for the cubic spline (green line) and arc-based (blue line) waveguide designs. Regions of small radii of curvature indicated (red circle).**

It is clear that the arc-based design has a constant radius of curvature of 23 mm along the entire length, except for a few points where the waveguide changes direction and where there is a tiny connecting straight section. On the other hand the cubic spline based waveguide has a more complicated radius of curvature profile: the majority of the waveguide length has a much larger radius of curvature than the arc-based design but with two small regions that possess lower radii of curvature, one of which is indicated by the red circle, where the radius of curvature drops as low as 11 mm in this region. This low radius of curvature is in the region below 23 mm in Fig. 7 that is to be avoided due to the substantial increase in measured bend loss over simulation.

Therefore, a waveguide with a more evenly distributed radius of curvature along the entire length will help to minimize bend losses by avoiding the low radius of curvature region. Indeed the predicted throughputs for the two curves in Fig. 10 are 37% and 54% for the cubic spline and arc designs respectively, based on a lookup table created from the measured power curve in Fig. 7. This highlights the fact that even small regions with very low radii of curvature can contribute significantly to the overall power throughput as illustrated earlier both analytically and via simulation. Thus, an outcome of this study is that all devices attempting to minimize bend losses should be designed with circular arcs. The circular arc design does have an increased transition loss over the cubic spline design, but is small compared to the beneficial decrease in bend loss. To mitigate the arc-to-arc transition, a bridging straight section could be used, but as we are in the low index contrast regime, the benefit of doing so would be minimal.

## 5. Designing an optimized, path length matched circuit

A new eight waveguide pupil-remapping device designed with arc-based curves was created and is depicted in Fig. 11. It can be seen that the new device has an additional feature: a lateral side step. In brief, this was necessary to avoid uncoupled light at the input of the waveguides spatially overlapping with the waveguides at the output and causing interference [24].

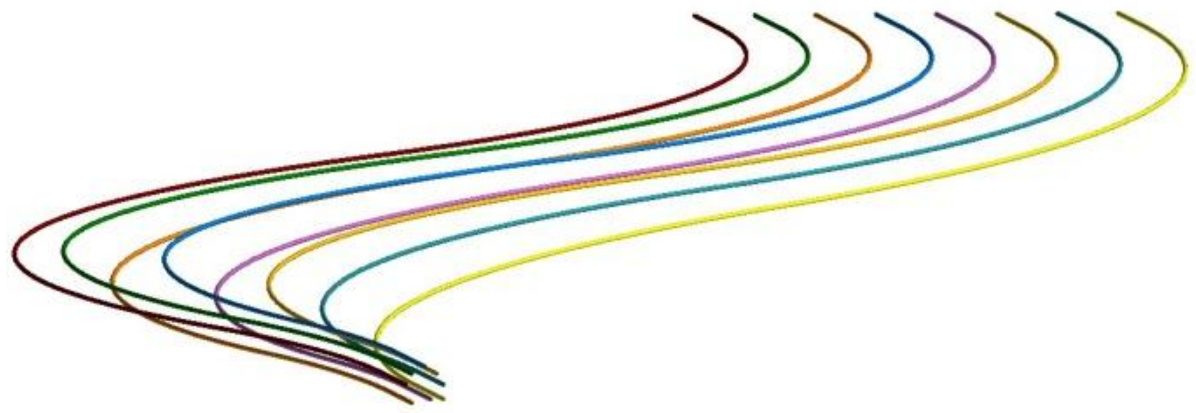

**Fig. 11 – Arc-based waveguide design of a pupil remapper with outputs shifted laterally by 5 mm at the output.**

The predicted throughput for the circuit in Fig. 11 was calculated and incorporates the bend losses from measured power in Fig. 7, the transition loss from Eq. 6 well, and the coupling and absorption losses previously mentioned. The predicted results for these eight waveguides are plotted in Fig. 12 along with the radius of curvature used in the design of each arc-based waveguide.

The radii of curvature are above 23 mm for all waveguides, which was the lower limit for the range over which the model could be trusted. The corresponding predicted throughputs were 37-59 ± 5%, which is sufficient for preliminary astronomical observations.

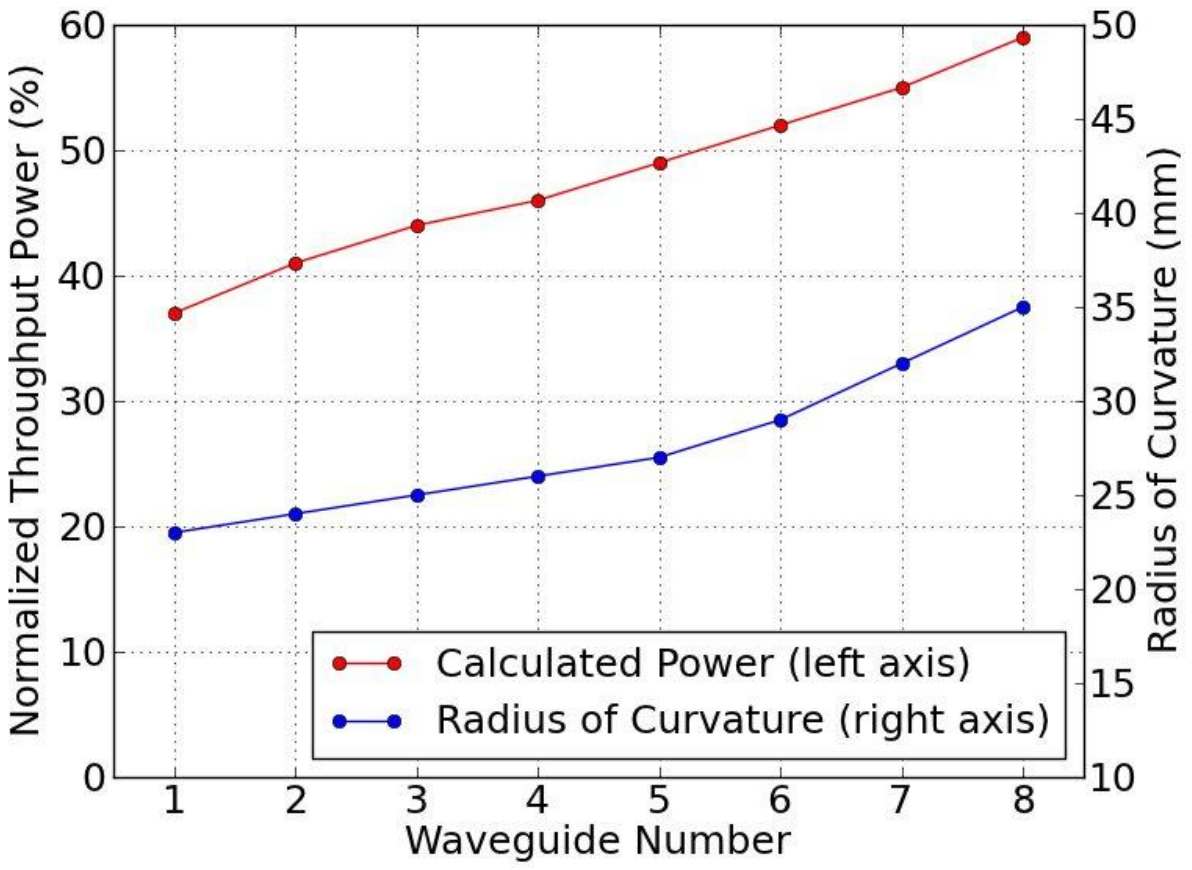

**Fig. 12 – Predicted throughput (red line) and corresponding radius of curvature (blue line) for each of the eight waveguides in Fig. 11.**

## 6. Device Fabrication and Characterization

The pupil remapper was inscribed using an ultrafast titanium sapphire oscillator (Femtolasers GmbH, FEMTOSOURCE XL 500, 800 nm centre wavelength, <50 fs pulse duration) with 5.1 MHz repetition rate. The laser was focused into a 30 mm long boro-aluminosilicate (Corning Eagle2000) glass sample using a 100× oil immersion objective lens (Zeiss N-Achroplan, numerical aperture (*NA*)=1.25, working distance = 450 μm). Pulse energies of 35 nJ were used in conjunction with translation velocities of 250 mm/minute in order to create waveguides which were single-mode at 1550 nm (see mode field and cross-sectional micrograph in Fig. 6(b)). A set of Aerotech, air-bearing translation stages was used to smoothly translate the sample during writing. The chip was ground and polished to reveal the waveguide ends.

To measure the accuracy with which the path lengths were matched in the pupil remapper, it was placed within a larger interferometric testbed (known as Dragonfly) [25]. A schematic diagram of the instrument is shown in Fig. 13. The testbed consisted of a segmented mirror (MEMS) that was reimaged downstream onto a micro-lens array, which was used to couple light into the eight waveguides simultaneously. By using this arrangement it was possible to fine-tune the coupling into each of the eight guides by steering the segments of the MEMs array. Indeed, by tilting the segments of the MEMs by a large amount the light could be completely decoupled from the guides. At the output the emerging beams were recollimated and then dispersed by a transmission grating onto an InGaAs sCMOS detector array.

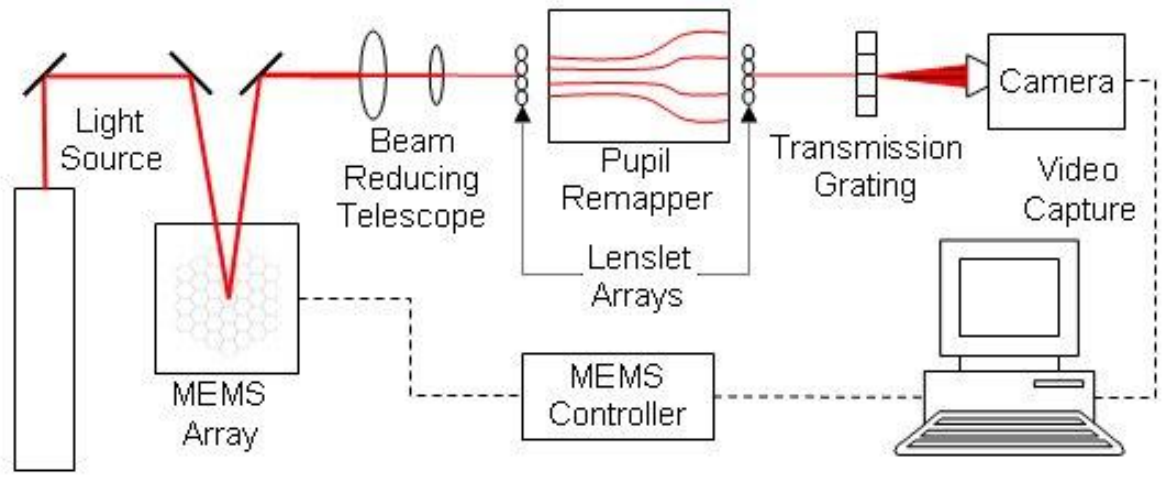

**Fig. 13 – Test setup for measuring the accuracy of path-length matching in the pupil remapping device.**

In order to measure the path length difference among all eight waveguides, one waveguide was selected as a reference and left on, while the other seven waveguides were switched on/off one at a time. This created two-beam interference patterns across all seven possible baselines, two of which are shown in Fig. 14.

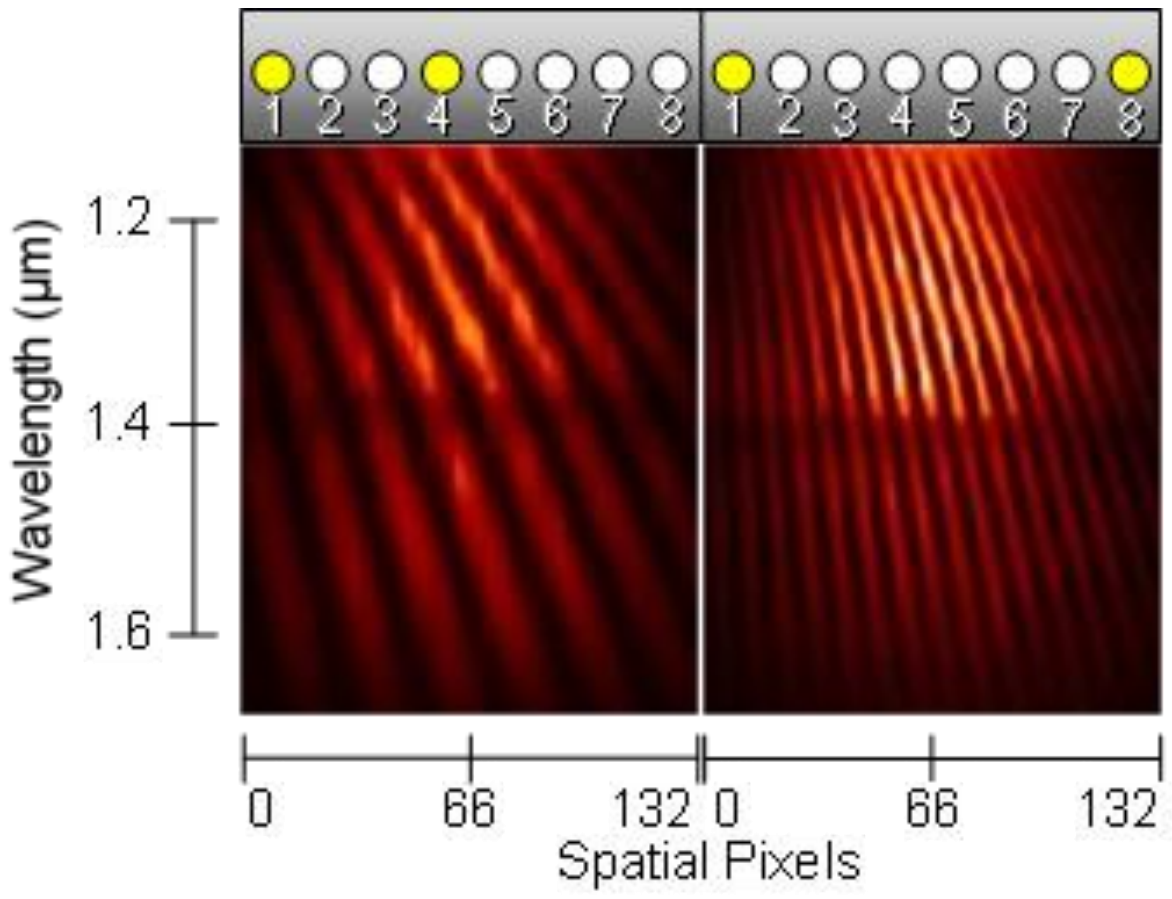

**Fig. 14 – Images of two sets of spectrally-dispersed interference fringes for two pairs of waveguides with smaller (left) and greater (right) separations. Wavelength runs in the vertical direction, while the spatial fringe pixels are horizontal. Top diagrams illustrate which two waveguides (yellow) are used to make the particular fringe image.**

By using the interferograms, it was possible to measure the phase delay between the wavefronts from the two waveguides generating the fringes. The vertical wavelength scale was first calibrated using measurements of narrowband laser sources at wavelengths within the fringe spectral range, which allowed the phase delay to be translated into a physical delay length. With the delay length measured for all seven waveguide pairs, the common reference waveguide's path length was set to zero. A graph of the relative path lengths of the seven waveguides is shown in Fig. 15.

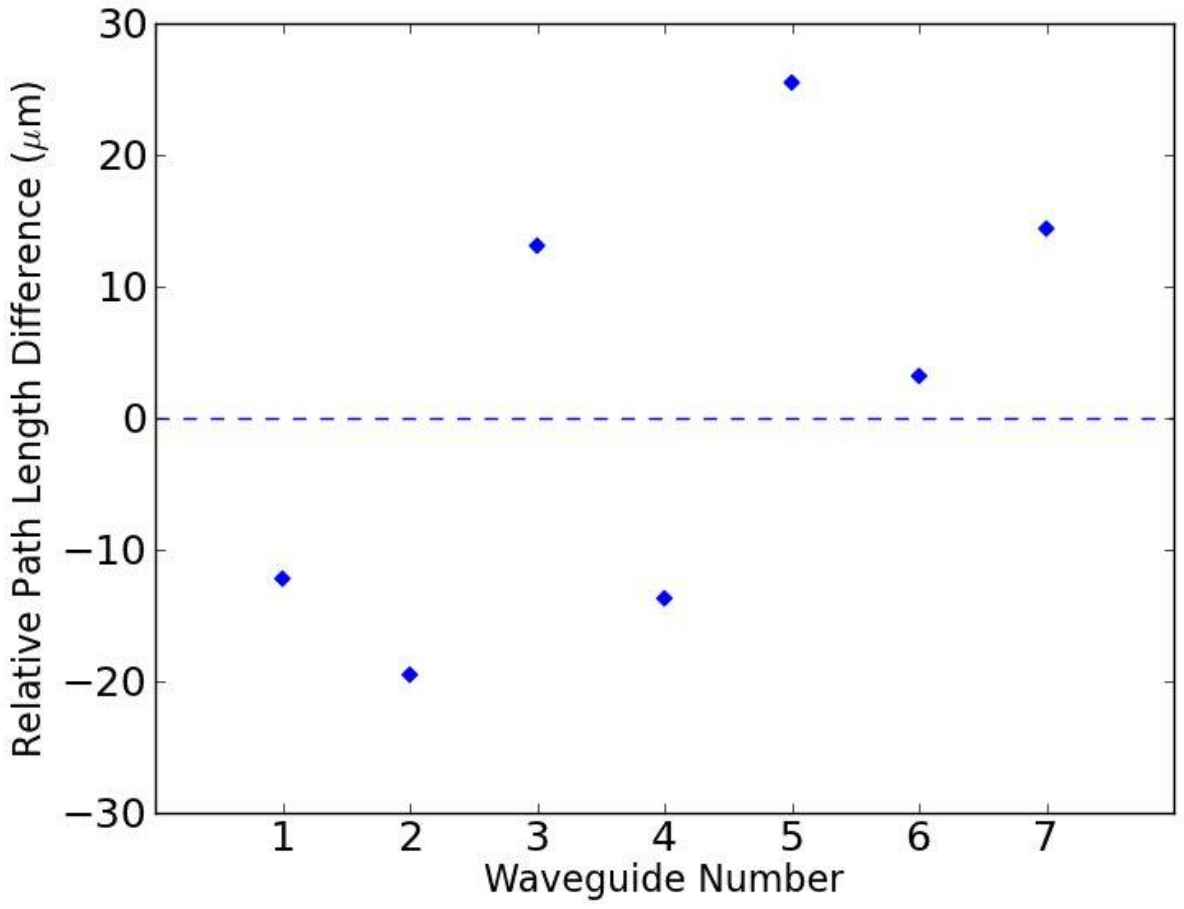

**Fig. 15 – Relative path length difference between each waveguide and the reference waveguide, which is normalized to zero.**

From the set of eight fabricated waveguides, the average path length difference of all possible waveguide pair combinations is 19.9 ± 5.8 µm with the longest path length difference between waveguide 2 (-19.5 µm) and 5 (25.5 µm) of 45 µm. As the coherence length of the light is given by $\lambda^2/\Delta\lambda$, where $\lambda$ is the wavelength of light and $\Delta\lambda$ is the bandwidth, then the maximum bandwidth that can be used while still maintaining high fringe visibilities is 54 nm (a 3.4% fractional bandwidth). This fractional bandwidth falls within the range typically used for astronomical observations and hence the device presented above is satisfactory for use in many observational contexts.

Although the performance reported here is within our design tolerance, there is a significant deviation between the ideal mathematical trajectories (which are designed for physical path length matching within 0.1 µm) and the measured path length matching of the fabricated component. Firstly, it is important to note that the path length differences reported above are not just for the chip in isolation but in fact are measurements of the entire interferometric testbed (from MEMS array to detector). This means that some of the path length mismatch may be due to the bulk optics and hence the values reported above are an upper limit for the chip in isolation. Secondly, the waveguides were designed to take into account physical path lengths by using mathematical trajectories and did not include any optical effects which could have caused the path length matching to deviate from the design. These include variations in the refractive index contrast between different guides (written at different depths with different curvatures), inhomogeneities in the index contrast along the guides due to laser power fluctuations or index-matching oil management issues, mode displacement around bends which affects the propagation velocity of the mode, polarization mode dispersion, and finally non-ideal performance of the stages causing errors in the guides. These errors are all very difficult to quantify. Although the design process does not take into account these optical affects, it is clear that the circuits are sufficiently well path length matched by just setting a condition for physical (rather than optical) path length matching. It would be possible to extend the design process to take into account some of these optical affects by way of simulations.

Finally, the transmission of all eight waveguides from the device fabricated based on the design of Fig. 11 were measured using the same probe technique described earlier, with resulting throughputs found to be between 5-47%. Although, these values did not match those predicted in Fig. 12, they were sufficient for initial testing. We have determined that the larger than expected losses are due to mechanical difficulties with our translation stages when moving at the high velocities used in our designs. We hope to overcome these fabrication issues in the near future and produce precisely path length matched, high throughput, complex three-dimensional photonic circuits.

## 7. Conclusion

We have outlined a method to design three-dimensional path length matched waveguides that take unique routes through a photonic device while simultaneously satisfying the numerous constraints of the device design and fabrication process. Examination of the radius of curvature and resulting bend loss illustrated the importance of minimizing curvature in waveguide design and found that arc-based guides are superior to splines in low-loss performance. We showed that beam propagation simulations were useful in predicting the performance of ULI waveguides. Finally, we fabricated and tested an eight waveguide path length matched pupil-remapping device and subsequently determined that all paths were matched to within 45 µm. Such performance is satisfactory for astronomical applications [26].